\documentclass{elsart}

\usepackage{harvard}

 \usepackage{graphics}
\usepackage{graphicx}
\usepackage{epsfig}

\usepackage{amssymb}

\def\url#1{{\ttfamily\def\/{/\discretionary{}{}{}}#1}}

\begin{document}

\begin{frontmatter}
\title{A universal density slope -- velocity anisotropy relation for 
relaxed structures}
\author{Steen H. Hansen, Ben Moore}
\address{University of Zurich, Winterthurerstrasse 190,
8057 Zurich, Switzerland}

\begin{abstract}
We identify a universal relation between the radial density slope
$\alpha (r)$ and the velocity anisotropy $\beta (r)$ for equilibrated
structures.  This relation holds for a variety of systems, including
disk galaxy mergers, spherical collapses, cold dark matter (CDM) halos
both with and without cooling.  We argue that the shape of the
relation is reasonable from fundamental principles when the dark
matter or stars are assumed to obey Tsallis statistics, and in that
case we fit the $\alpha - \beta$ relation with just one free
parameter. One can use this result to close the Jeans equations, for
example to construct mass models of elliptical galaxies from
observational data or to tune dark matter direct detection
experiments. We also predict the asymptotic central slope and
anisotropy of CDM halos to be approximately $-1$ and $0$.
\end{abstract}

\begin{keyword}
\end{keyword}
\end{frontmatter}


\section{Introduction}

Gravitating equilibrium structures must satisfy the Jeans equations, however
these are generally difficult to solve - especially for multi-component or
flattened systems. One of the main difficulties is that 
there is no clear connection between the density and
the components of the velocity dispersion tensor. Even for the simplest
gravitationally bound systems of collisionless particles one must make
strong assumptions to find a solution. One example hereof is the
central density profile of dark matter structures, which has still not
been derived directly from the collisionless Boltzmann equation. Instead one must
resort to N-body simulations which have been used to demonstrate
a universality amongst the density profiles of CDM structures,
see e.g.~\cite{diemand}, and references within.
Only when making strong assumptions about a connection between the
density and the velocity dispersion can one find analytical 
solutions~\cite{binneytremaine}.  
E.g. under the assumption that the phase-space
density is a power-law in radius can one solve for the density 
profile~\cite{hansen,austin05,dehnen05}.
Furthermore, if one assumes a given form for the density profile and
assumes a given radial dependence of the anisotropy, then one can
derive the radial behaviour of mass, potential and velocity
dispersion, see e.g.~\cite{lokasmamon}
for references. 
Similarly, it has also been suggested that the
velocity anisotropy should follow a universal profile in
radius for dark matter structures~\cite{cole,carlberg}, however these 
relations are not generally valid, as will be shown in Fig.~1.

We will here address a related question, namely whether there
is a connection between the radial derivative of the density profile,
the density slope $\alpha(r)$, 
\begin{equation}
\alpha (r) \equiv \frac{d {\rm ln} \rho}{d {\rm ln} r} \, ,
\end{equation}
and the velocity anisotropy $\beta(r)$
\begin{equation}
\beta (r) \equiv 1 - \frac{\sigma^2_t}{\sigma^2_r} \, ,
\end{equation}
where $\rho (r)$ is the density, and $\sigma^2_t$ and $\sigma^2_r$ are 
the one dimensional tangential and radial velocity dispersions. 
We will show that this connection is
universal (within the scatter from the numerical simulations),
and we will argue that such a connection is reasonable from a
fundamental statistical mechanics point of view. We find that the $\alpha -
\beta$ relation can be argued (or maybe eventually derived) from first
principles, with only one free parameter.

\begin{figure}
\begin{center}
\epsfxsize=8.5cm
\epsfysize=6.5cm
\epsffile{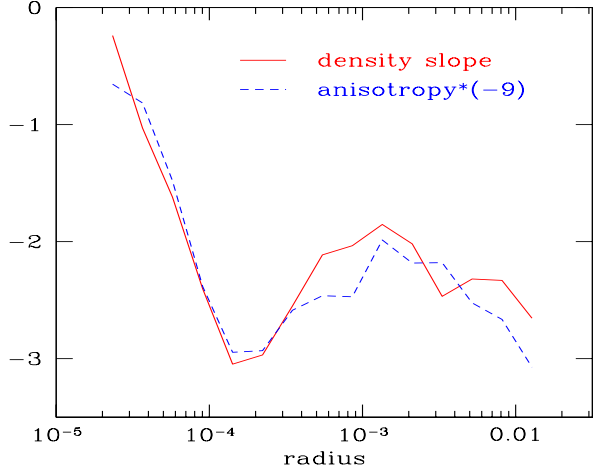}
\end{center}
\caption{The radial anisotropy and density profile slope of an 
equilibrated dark matter structure where $5\%$ of the particles are
allowed to cool in order to mimic baryons. The red (full) line is the
density slope of all gravitating matter, and the blue (dashed) line is
a simple linear function of the DM anisotropy.}
\label{fig1}
\end{figure}

\section{A relation between $\alpha$ and $\beta$}

In order to address the question of a possible relation between the
density slope $\alpha$ and the velocity anisotropy $\beta$ we first
consider
a result from an N-body simulation. This simulation follows the hierarchical formation of
a dark matter structure where $5\%$ of the particles are allowed to
cool, mimicking the behaviour of baryons. The reason for considering
this specific simulation is that it is remarkably complex and 
captures many of the physical processes that may arise during
the formation of galactic systems. The final dark matter structure
is built up of many hundreds of minor and major mergers of 
small halos. The resulting triaxial structure is modified 
by the effects of dissipating particles, deepening the potential,
altering the orbital classes of dark matter particles and 
changing the global shape of the system. The final system
does not follow the universal density profile and has a complex
distribution of dark matter and particles allowed to cool - the latter
dominate the potential within a few percent of the virial radius.  

The question of resolved region and whether the
structure at a given (large) radius is equilibrated will be discussed later.
The result is presented in Fig.~1, 
where the red (full) line is the density slope of all the matter
present. On the same figure we present a simple linear function of the
anisotropy, $ -9 \cdot \beta$, 
as a blue (dashed) line.
This demonstrates
a clear connection between the density slope and the anisotropy. If
a relation reveals itself in this highly complex simulation, then we
expect that a similar relation might hold for {\it any} relaxed and
equilibrated structure.

It is important to clarify that whereas universality has previously
been suggested both for the radial density profile~\cite{nfw,moore}
and the radial anisotropy profile~\cite{cole,carlberg}, then the
simulated density and anisotropy profiles presented in Fig.~1 clearly
do not follow such universal behaviour: this non-universality at
intermediate radii is seen as a dip in Fig.~1, which is absent in the
previously suggested universal profiles.
Nevertheless, the anisotropy and density slope have a
clear correlation for all radii.

When we discuss relaxed structures, we mean that the system has
undergone rapid processes of violent restructuring to attain a 
new quasi-equilibrium state, such as a merger,
i.e. an elliptical galaxy resulting from the merger between two disk galaxies. 
A true equilibrium never occurs for any
astrophysical system since they are not ideal fluids, rather they
are made of massive objects such as stars (or dark matter particles)
which undergo continuous short and long range interactions. Over very
long timescales this collisionality may cause the central
regions to collapse and the outer particles to disperse. For a spiral
galaxy, the stars will evolve from a disk into a spheroidal
structure over a timescale of $\sim 10-100$ Hubble times.

\section{A universal $\alpha - \beta$ relation}

Let us now explore the properties of four additional different simulations.

(i) Firstly we consider a spherical collapse calculation. We place one million
particles in a spherical distribution with an initial $r^{-1}$ density
profile. The system has unit mass an initial unit size and the particles
have zero velocity. We run the simulation
for over 10 dynamical times $t_{dyn} \propto {(G\rho)}^{-0.5}$.
The system maintains its symmetry only until the first shells cross 
in the central region. The radial orbit instability occurs
and the infalling shells of particles undergo strong scattering 
from a central bar-like structure. The final equilibrium structure
has a half mass radius of about 0.1 and is prolate with
axial ratios of about 2:1. The density profile has a slope of -1.3
in the innermost resolved region - similar to that found within 
the cosmological simulations of CDM halos.

(ii) Our second study starts from two identical equilibrium 
NFW structures each with zero anisotropy. These are constructed 
using the technique described in \cite{stelios}. 
Each halo has one million particles and has a mass of $10^{12}M_\odot$
and a concentration parameter $c_{\rm nfw}=10$. We carry out two
simulations in which the halos collide head on producing a prolate
structure, and a merger with angular momentum which produces
an oblate structure. The simulations are run for 10 Gyrs to allow
the merger remnant to attain equilibrium,
see~\cite{moore2004} for further details.

(iii) We carry out a merger between two disk galaxies that
are constructed to be similar to a late type disk galaxy such
as M32. The halo is initially NFW but adiabatically contracted 
due to the presence of an exponential disk of stars. The two
disks are placed on a parabolic inclined orbit typical of 
those found within cosmological simulations. We analyse
the simulation after 5 Gyrs after which the system resembles
a relaxed elliptical galaxy.

(iv) Finally we take a high resolution cosmological simulation of a
cold dark matter halo which is resolved with over ten million particles.
This is one of the cluster mass halos from the study 
of \cite{diemand}.
The density profile attains a slope of -1.3 in the innermost resolved
region.

For each of these simulations, which have been allowed to
approximately equilibrate, we extract the density profile and the
radial dependence of the anisotropy. The results are shown in Fig.~2
as symbols.
\begin{figure}
\begin{center}
\epsfxsize=8.5cm
\epsfysize=6.5cm
\epsffile{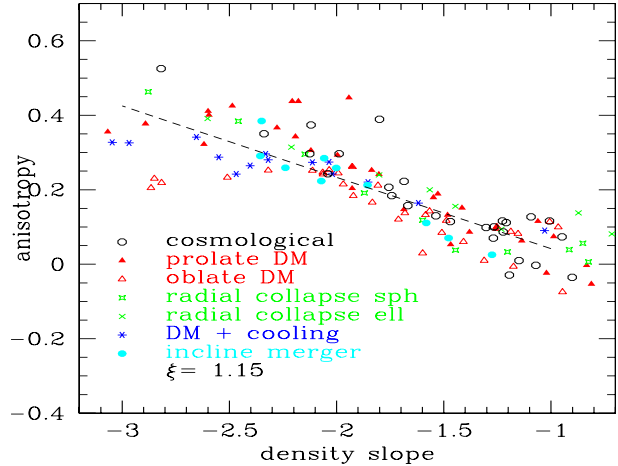}
\end{center}
\caption{The final density slope versus velocity anisotropy for the
six different simulations (coloured symbols) discussed in the text. 
For the spherical collapse calculation we plot the result using both
spherical and elliptical bins.
A general relation is clear with the slope in the range
-3 to -1.  The  shape of
the relation is reasonably well fit with 2 parameters, $\beta =
\eta_1 - \eta_2 \, \alpha$, where approximately $ -0.45 < \eta_1 < 0.05$
and $ 0.1 < \eta _2 < 0.35$.  
The dashed (black) line is a one parameter fit as discussed
in section~\ref{sec:discussion}, with the parameter $\xi = 1.15$.}
\label{fig2}
\end{figure}
Surprisingly, there seems to be a clear trend, namely that for
these completely different simulations the relation between the
slope of the mass profile 
and the anisotropy parameter of the particles
is the same.  The simulations are roughly
resolved to the radius where the slope is approximately $-1$, 
and should not be trusted for more shallow profiles. This resolution
limit is set by the number of particles in the central regions - we 
do not know if the correlation we find is affected by the numerical
relaxation processes that occur where particle-particle encounters
are important.
For slopes steeper than $\approx -2.5$ the
particles have often not had time to equilibrate fully. 

In principle, the anisotropy parameter could have any value
in the range $-\infty
< \beta \leq 1$. Isotropic orbits have $\beta=0$ and positive
values correspond to more radial orbits.
On Fig.~2 we see that the relations is approximately fit with
$\beta = \eta_1 - \eta_2 \, \alpha$.
More detailed simulations and convergence tests are in progress
to clarify the best fitting relation and to investigate
the cause of the scatter. It will be interesting to explore
how much ``violent relaxation'' is required to force an
existing equilibrium structure to lie on this correlation, and what
the driving force behind
establishing the correlation is. 
The time evolution of the radial infall simulation allows us to 
compare the change of anisotropy with the change in the mass profile.
Even though naively, one
might have expected that the density profile should first manifest
itself and then only later the anisotropy should appear, it
seems that the two come hand in hand. As soon as a given density 
profile exists (and is approximately equilibrated) then the
anisotropy exists as well, and sits near the $\alpha - \beta$ relation.
We also point out that the orbital structure of the models
does not play an important role in changing the observed correlation.
The halo merger simulations produced a prolate halo with over 90 percent
of the particles on box orbits whilst the oblate halo was primarily
tube and loop orbits, yet both models lie on the $\alpha -\beta$ relation.

\section{Discussion}
\label{sec:discussion}

The relation identified in the previous sections turns out to be
rather non-trivial to explain, and the intuitive arguments of this
section will make it evident that there is still much work to be done.

For a classical gas the velocity dispersion is like a temperature. If
you have a box of classical particles then the velocity dispersion
seen from one side of the box will be the same as seen from another
side of the box. For the dark matter structures the anisotropy is
therefore like having {\em different} temperatures when considering
the radial and tangential directions. The question we are posing is,
why is this related to the density profile? Since these completely
different simulations described above seem to give the same relation,
then the reason must be of fundamental nature. Let us therefore
consider the velocity distribution function, $f(v)$.

The existence of a non-zero anisotropy implies that the velocity
distribution function (VDF) in the radial direction, $f_r$, differs
from the VDF in the tangential direction, $f_t$.  Intuitively one may
in fact expect $f_r$ and $f_t$ to differ, simply because the structure
is in equilibrium which implies that the VDF's are constant in shape
and magnitude. That is seen as follows: particles moving tangentially
experience no change in energy, so the conservation of $f_t$ is
trivial and may be obtained by virtually any shape of $f_t$. The height
of $f_t$ is fixed by the density which is constant along the
tangential direction. Contrariwise, particles moving radially will
both be changing kinetic energy as well as moving to a region with a different
density, thus both the shape and magnitude of $f_r$ is non-trivial to
conserve. This puts strong constraints on the shape of the radial
VDF. We thus see that there are strong constraints on the shape of
$f_r$, and virtually no constraints on $f_t$. There is therefore very
little reason to expect isotropy.  Now, let us quantify these
statements.

It is well known that for a given radial density profile (and zero
anisotropy)
one can use the Eddington inversion method to extract the velocity
distribution function, $f(v)$, at different radial bins. The shape
of this distribution function can be obtained analytically for sufficiently
simple structures, and in particular for isotropic structures with constant
density slope the shape is the Tsallis distribution function, $f(v,q)$, 
which depends on the non-extensive index $q$~\cite{tsallis88,plastino93}.
In fact, a simple connection between this non-extensive parameter, $q$,
and the density slope has been derived by \cite{hansen04}
\begin{equation}
\alpha =  \frac{6q - 10}{3 -q} \, ,
\label{eq:aq}
\end{equation}
where typically $0< q < 5/3$.  
This equation strictly speaking only holds for isotropic structures.
However, the Eddington inversion used to derive this equation is only
referring to the radial component of the velocity distribution and
simply assumes that the tangential component is identical to the
radial component. We will now use eq. (3) to argue that there really
should be a difference between the radial and tangential distribution
functions, and that real structures therefore should be anisotropic.
To test the accuracy of our arguments below, one should derive a
connection similar to eq. (3) for anisotropic structures, and if that
connection would be similar to eq. (3), then our arguments are
reliable.

From the expression in eq.~(\ref{eq:aq})
it is clear that for constant density ($\alpha=0$) one
has $q=5/3$. Thus, the tangential velocity distribution function will
have a shape given by $q=5/3$, whereas the radial velocity
distribution function will generally have a shape given by a
different $q$.
We therefore understand that the particle
distributions generally are different in the radial and the tangential
directions. It is reasonable to speculate that it really is the
changing potential energy being the driving force between the difference
in the radial and tangential direction.

Recently it has indeed been confirmed that the velocity distribution
function for the radial and tangential part indeed can be described by
using different $q$'s as suggested above~\cite{veldistr}.
In particular was it shown that the bulk of the
tangential distribution function indeed should be described by $q=5/3$
as we have just argued. This is strong support that our arguments
presented above are fairly reliable.

One can directly derive the fraction between 
velocity dispersions for different $q$'s. When the energy of the particles
is trivially given as $E \sim v^2$, that is, if the potential energy was 
constant, then the result is
\begin{equation}
\frac{\sigma^2_t}{\sigma^2_r(q)} =  \frac{S(t)}{S(r)}  \, \frac{A(t)}{\zeta(t)} \, \frac{\zeta(r)}{A(r)} \, ,
\end{equation}
with
\begin{equation}
S(q,D) = \frac{2D}{(D+2) - Dq} \, , \label{eq:sterm}
\end{equation}
where $\zeta$ is the Lagrange multiplier and should cancel,
$A= 1 + (1-q)D/2$~\cite{boghosian}, 
and we have explicitely written the $r$ and $t$ for radial
and tangential. 
These equations only hold when the distributions, radial and
tangential separately, can be approximated by the Tsallis shape.
The S-term was previously derived for 3
dimensions~\cite{silvaalcaniz},
and the q-dependence of A may be important for cluster temperature
determination~\cite{cluster}. Using eq.~(\ref{eq:aq}) we have
\begin{equation}
A(\alpha, D) = \frac{\alpha \left( 1 - D \right) + \left(
6 - 2 D\right) }{6 + \alpha} \, .
\end{equation}
Since we are considering 1-dimensional velocity
dispersions we have $D=1$ and one finds
\begin{equation}
\frac{A(t)}{A(r)} = \frac{6+\alpha}{6} \, .
\label{eq:aa}
\end{equation}

In the simplest of all worlds one would indeed have $E \sim v^2$, in
which case the S-term in eq.~(\ref{eq:sterm}) 
would exactly cancel with eq.~(\ref{eq:aa}),
and we would have identical zero anisotropy everywhere.
One can, however, imagine more complicated situations where this does
not happen since the potential energy changes as function of radius, 
so that the S-term would look different. 
Phenomenologically we parameterise our ignorance with an unknown
constant, $\xi$, and we therefore conclude that the anisotropy may look like
\begin{equation}
\beta(\alpha) = 1 - \xi \, (1 + \alpha/6)  \, .
\label{eq:fit}
\end{equation}
The points on Fig.~2 are approximated with $\xi \approx
1.15$, as shown with the dashed (black) line.

The first obvious objection is that whereas a connection between the
slope and the anisotropy is clear from Fig.~1, there is a large
scatter in Fig.~2. It may in fact be that the relation is not
universal after all, in which case it would make no sense trying to
derive it from first principles.  To this end one must keep in mind the
large range that our simulations span. They include both cosmological
simulations and also controlled colliding initially isotropic 
models ($\beta=0$) and radial collapse with $\beta=1$.
It would be an impressive coincidence if these
simulations would conspire to accidentally land {\em near} the same
line in $\alpha - \beta$ space. At a given value of the density profile
the scatter in the anisotropy parameter is about $0.2$.

The second objection is that it is possible to create a structure
e.g. with a Hernquist density profile and with zero anisotropy everywhere,
which is stable, so the suggested $\alpha - \beta$ relation cannot
be a necessary criteria for all equilibrium systems.
This is similar to the statement that it is
possible to arrange a configuration of classical particles in a box
which all move exactly in the {\it x}-direction only. This configuration is
in principle also stable, however, any small perturbation
(e.g. inclusion of gravity) will perturb the system, and soon the
classical particles will achieve an isotropic distribution.
Similarly, the suggested Hernquist structure may (and we
suggest
 \footnote{ We performed a controlled experiment, where each
individual particles in an initially stable Herquist structure is
perturbed slightly and then the system is allowed to equilibrate,
repeatedly.  These perturbations thus act like a non-zero term on the
r.h.s.  of the boltzmann equation, and hence antonov's laws of
stability may not apply.  We indeed observe that the structure slowly
moves towards the $\alpha-\beta$ line.  } that it will) move away from
the initial zero anisotropy, and flow towards the line in
$\alpha-\beta$ space, when the system is exposed to significant
perturbations, such as merging of substructures.  A large perturbation
of the artificially constructed classical box of particles will
rapidly lead to equilibration, whereas a small perturbation will only
slowly lead to equilibration.  Similarly, we expect that a larger
perturbation will drive any equilibrium structure towards the
universal $\alpha-\beta$ line.

The third objection is our argued fitting formula,
eq.~(\ref{eq:fit}). This must clearly break down for small density
slopes.  
The reason is that when the slope is zero, then there is basically no
difference (from the point of view of the velocity distribution
function as explained above) of going in the radial or the tangential
direction, and hence the anisotropy may go to zero.
Furthermore,  eq.~(\ref{eq:aq}) was derived for
isotropic structures. It may, however, be that the correct relation
between $\alpha$ and $q$ should depend on the anisotropy as well.  In
that case the arguments above will change.

We note that our finding has consequences for dark matter detection,
since the Earth is expected to be near $\alpha \approx -1.3$ which implies
$\beta \approx 0.1$.
First the existence of an anisotropy will alter the
annual modulation~\cite{vergados,evans00}, and second the anisotropy  
will also allow for a 'non-annual-modulation'
differentiation between signals tangentially and radially w.r.t.\ the
galaxy center with a directional sensitive DM detector.

We can finally combine our findings with known results, in order to
predict the central slope and anisotropy of equilibrated dark matter
structures. 
One can find a very general bound on the central anisotropy as
function of the central density slope, $\beta \leq \gamma/2$, which is
valid for all systems obeying the Jeans equation \cite{An.Evans}. Much
stronger bounds naturally appear when considering more restrictive
systems. When assuming that phase-space density is a pure power-law in
radius (as observed in systems from cosmological simulations
\cite{TaylorNavarro}), then the bound on the central slope as a
function of the anisotropy is instead~\cite{hansen}
\begin{equation}
\alpha (0)  \leq  - \left( 1 + \beta(0) \right) \, , 
\end{equation}
which together with the $\alpha - \beta$ relation implies that
the central values are given by 
$ ( \alpha (0), \beta (0) )  \approx   (-1, 0)$. 
One should keep in mind the scatter in Fig.~2, which implies
that this bound still has significant 
uncertainties~\footnote{Recently Dehnen \& McLaughlin 
(2005) refined the analysis of Hansen
(2004), and they find the central slope to be connected to the central
anisotropy as $\gamma_0 = -(7 + 10\beta_0)/9$, also under the
assumption that phase-space density is a power-law in radius as
observed in N-body simulations.  When using this analytical result
together with the numerical findings of this paper do we conclude that
the central anisotropy is close to zero, and the central density slope
is close to $-0.8$.}.

\section{Conclusions}
\label{sec:con}
We have identified a relation between the density slope and the
velocity dispersion for dark matter structures, and we suggest that
this relation may be universal. We support this with a set of
completely different simulations, which all land near this $\alpha
-\beta$ relation. We argue that the relation is reasonable from a
fundamental statistical mechanics point of view, when the star or dark matter
distribution functions are to be described by Tsallis statistics, and
the relation is in that case fit with just one free parameter. We use
our result to predict that the central slope and velocity anisotropy
are approximately $(\alpha(0), \beta(0)) \approx (-1, 0)$.

This remarkable correlation implies that in all systems that have 
undergone a violent relaxation process, the anisotropies of all
components (e.g. stars and dark matter) are identical to each other.
Utilising this fact allows a unique solution to the Jeans equation
to be established for any given system. No longer is the anisotropy
a free parameter - it can be related to the local slope of the
total density profile. This will allow unique mass models to be
constructed for elliptical galaxies for example. Given good enough
data, one could measure the anisotropy of the stars in such a system
to infer the structure of the dark matter halo. It can also be used
to tune direct detection experiments.

\section*{Acknowledgements}
It is a pleasure to thank Juerg Diemand and Joachim Stadel
for discussions and simulations.
We thank Wyn Evans for insightful comments.
SHH thanks the 
Tomal\-la foundation for financial support.

\end{document}